\documentclass[aps,showpacs,prx,amssymb,amsmath,twocolumn]{revtex4-2}
\usepackage{graphicx}
\usepackage{amssymb}
\usepackage{amsmath}
\usepackage{amsthm}
\usepackage{bm}
\usepackage{xcolor} 
\usepackage[dvipsnames]{xcolor} 
\usepackage[x11names]{xcolor}
\usepackage{array}
\usepackage{multirow}
\usepackage{tabularx}
\usepackage{booktabs}
\usepackage{textcomp}
\usepackage{mathtools}
\usepackage{hyperref}
\usepackage{qcircuit}
\usepackage{bbm}
\usepackage{cancel}
\usepackage{comment}
\usepackage{physics}
\usepackage{romannum}
\usepackage{kotex}
\usepackage{appendix}
\usepackage{cleveref}
\usepackage{orcidlink}
\crefname{equation}{Eq.}{Eqs.} 
\AtBeginDocument{\pagenumbering{arabic}}

\begin{document}

\title{Enhanced Maximum Independent Set Preparation with Rydberg Atoms Guided by the Spectral Gap}
\author{Seokho Jeong\,\orcidlink{0000-0002-7219-8955} and Minhyuk Kim\,\orcidlink{0000-0001-8705-7795}} \email{minhyukkim@korea.ac.kr}
\address{Department of Physics, Korea University, Seoul 02841, Republic of Korea}
\date{\today}

\begin{abstract} \noindent

Adiabatic quantum computation with Rydberg atoms provides a natural route for solving combinatorial optimization problems such as the maximum independent set (MIS). However, its performance is fundamentally limited by the reduction of the spectral gap with increasing system size and connectivity, which induces population leakage from the ground state during finite-time evolution. Here we introduce the Adjusted Detuning for Ground-Energy Leakage Blockade (ADGLB), a spectral-gap-guided schedule engineering method that modifies the laser detuning profile to suppress leakage without introducing additional Hamiltonian terms or iterative optimization loops. We experimentally benchmark ADGLB on a quasi-one-dimensional chain of $N=10$ atoms, and the MIS preparation probability increases substantially compared with the standard adiabatic schedule. Furthermore, we show that the schedule optimized for smaller instances can be directly applied to larger two-dimensional triangular lattices with $N=25$ and $N=37$. With a small heuristic offset, the method also remains effective for instances with higher hardness parameters. These findings demonstrate that spectral-gap-guided schedule engineering offers a scalable and hardware-efficient strategy for enhancing adiabatic quantum optimization on neutral-atom platforms.

\end{abstract}

\maketitle

\section{Introduction} \noindent
Over the past few decades, advances in the experimental control of quantum systems have brought quantum information processing closer to practical relevance~\cite{Feynman_IJTP_1982, Nielsen_Chuang_Book_2010, Dowling_PTRSA_2003, Shor_IEEE_1994, Grover_PRL_1997}. Both analog and digital approaches to quantum evolution have been actively explored, enabling applications in combinatorial optimization~\cite{Farhi_Science_2001, Dickson_PRL_2011,VickyChoi_Arxiv_2010, Niroula_SD_2022,Zhu_QST_2023, Dupont_SciAdv_2023,Dupont_PRAppl_2025, JSH_PRR_2023,Ebadi_Science_2022}, quantum simulation of many-body physics~\cite{Ebadi_Nature_2021,Leseleuc_Science_2019, Zhang_Science_2023,Fauseweh_NatComms_2024, Monroe_RMP_2021,Kaplan_PRL_2020,Wang_PRA_2012,Pagano_QST_2019}, and quantum error correction~\cite{Pichler_PRL_2025, Satzinger_Science_2021,Acharya_Nature_2025,Bluvstein_Nature_2026, RodriguezBlanco_PRA_2024,Butt_NatComms_2026}. These developments have been implemented across promising platforms such as trapped ions~\cite{Monroe_RMP_2021, Niroula_SD_2022, Zhu_QST_2023, Kaplan_PRL_2020, Wang_PRA_2012, Pagano_QST_2019, RodriguezBlanco_PRA_2024, Butt_NatComms_2026}, superconducting circuits~\cite{Arute_Nature_2019,Dupont_SciAdv_2023,Dupont_PRAppl_2025,Zhang_Science_2023,Fauseweh_NatComms_2024,Satzinger_Science_2021,Acharya_Nature_2025}, and neutral atoms~\cite{Pichler_PRL_2025,Ebadi_Nature_2021,Ebadi_Nature_2021,JSH_PRR_2023,Ebadi_Science_2022,Leseleuc_Science_2019,Bluvstein_Nature_2026}.

Among these platforms, Rydberg-atom arrays combine flexible programmability with intrinsic scalability~\cite{Saffman_RMP_2010}. Strong, distance-dependent Rydberg interactions naturally induce entanglement between nearby atoms~\cite{Gallagher_Book_1994, Endres_Science_2016}, while site-resolved laser control enables the implementation of tunable many-body Hamiltonians and programmable unitary dynamics~\cite{Adams_JPhysB_2019, Saffman_JPhysB_2016}. These capabilities have established Rydberg systems as a versatile platform for quantum optimization and simulation~\cite{Adams_JPhysB_2019, Saffman_JPhysB_2016, Barik_FQST_2024, Morgado_AVSQuantSci_2021}.

The maximum independent set (MIS) problem provides a particularly accessible application of adiabatic quantum computing (AQC) in Rydberg-atom platforms, offering a physically intuitive scheme in which solutions are encoded in the ground state of a gradually evolving Hamiltonian~\cite{Farhi_Science_2001, Dickson_PRL_2011}. This problem can be encoded naturally using Rydberg blockade, where constraints arise directly from interaction-induced energy penalties~\cite{JSH_PRR_2023, MHKim_NPhys_2022, Pichler_Arxiv_2018, Ebadi_Science_2022}. Recent experiments have demonstrated MIS on instances with various connectivity structures and extended this approach to broader classes of optimization problems through MIS mappings~\cite{Pichler_Arxiv_2018, Ebadi_Science_2022, MHKim_NPhys_2022, Dalyac_PRA_2023, AByun_PRXQ_2022, KKH_SD_2024, PJY_PRR_2024, YSong_PRR_2021, AByun_AQT_2024, Oliveira_PRXQ_2025, JSH_PRR_2023, JSH_AQT_2025}.

Despite these advances, the performance of AQC remains fundamentally constrained. In particular, the minimum spectral gap typically decreases with increasing system size and depends sensitively on the problem instance, thereby leading to population leakage from the ground state for a finite evolution time~\cite{Jansen_JMPhys_2007, Braida_Quantum_2025, Mozgunov_PTRSA_2022}. Consequently, designing the adiabatic evolution under a fixed total evolution time is crucial. This limitation is closely related to the instantaneous spectral properties of the system; however, accessing this information for large-scale quantum Hamiltonians remains challenging.

Alternative approaches include analog counterdiabatic quantum computing~\cite{Zhang_Arxiv_2024, Hsieh_Arxiv_2025, Perseguers_PRApplied_2025}, which typically requires additional Hamiltonian terms. Variational algorithms and Bayesian optimization~\cite{Finzgar_PRR_2024} rely on repeated circuit executions and extensive sampling, leading to substantial hardware overhead. Classical post-processing and mitigation techniques can partially improve performance, but their effectiveness ultimately depends on the quality of the probability distributions produced by the quantum hardware.

\begin{figure*}[thb!]
    \centering
\includegraphics[width=1\textwidth]{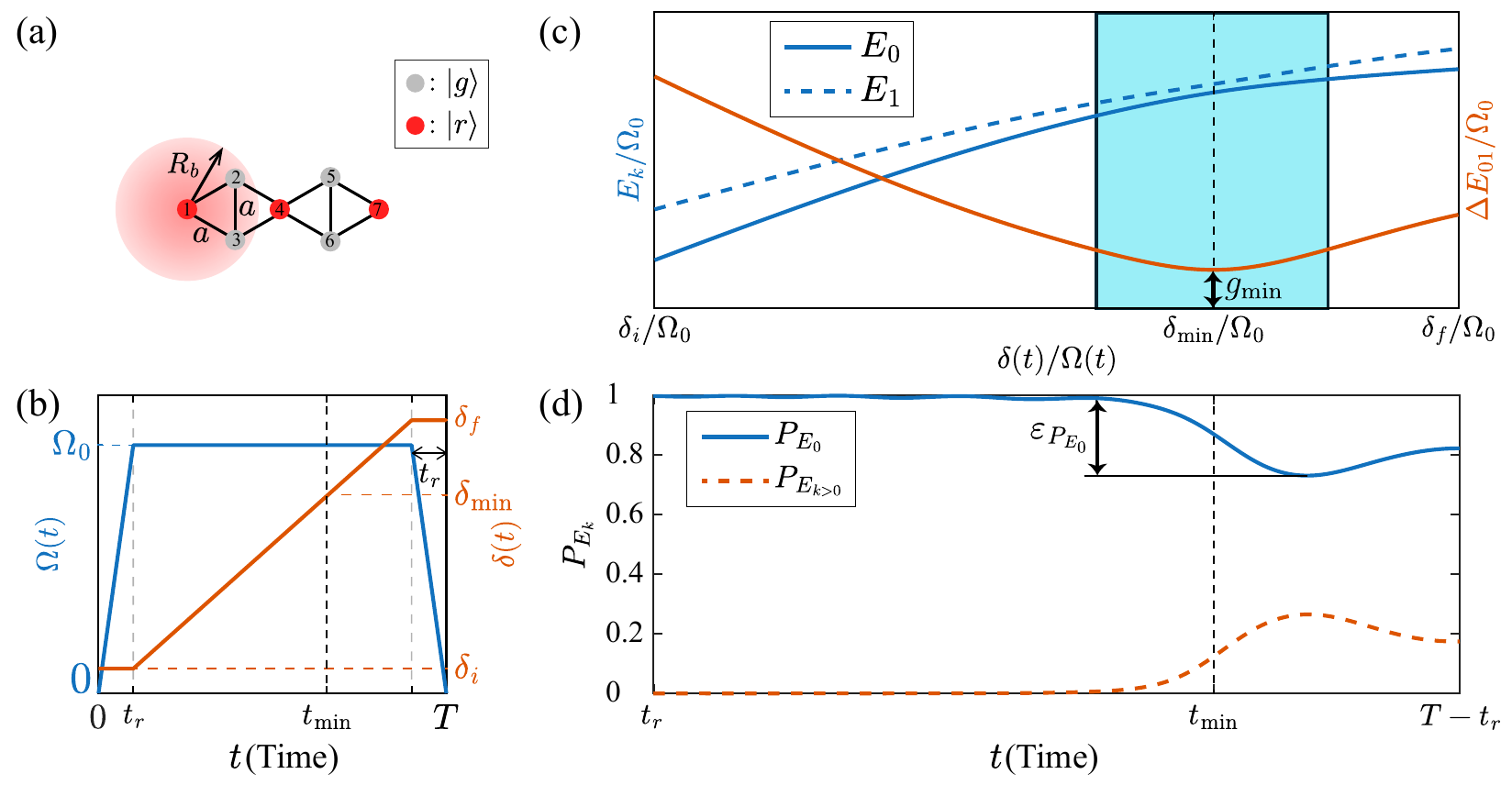}
    \caption{AQC for MIS preparation with Rydberg atoms. (a) An example of a MIS encoded in a $N=7$ Rydberg-atom array with nearest-neighbor spacing $a$, where atoms within the Rydberg blockade radius $R_b$ cannot be simultaneously excited. The atomic ground state $\ket{g}$ and the Rydberg state $\ket{r}$ are represented by gray and red circles, respectively. (b) The standard AQC schedule of the Rabi frequency $\Omega (t)$ and the detuning $\delta (t)$. (c) The ground-state energy $E_0$ (blue solid line), the first excited-state energy $E_1$ (blue dashed line), and the spectral gap $\Delta E_{01}\equiv E_1 - E_0$ (red solid line) as a function of the detuning $\delta (t)$ during the linear sweep $(t_r\leq t \leq T-t_r)$ of the standard schedule shown in (b). The region around the minimum spectral gap $g_{\rm min} \equiv \min_{t \in [0, T]} \Delta E_{01}(t)$ near $\delta_{\text{min}}$ is highlighted in light blue. (d) The time evolution of the ground-state population $P_{E_0}$ (blue solid line) and the population of all excited states $P_{E_{k>0}} \equiv 1-P_{E_0}$ (red dashed line). The population leakage from the ground state is denoted by $\varepsilon_{P_{E_0}}$.
}
\label{Figure1}
\end{figure*}

In this work, we demonstrate an enhancement in MIS preparation guided by the spectral gap. We introduce the Adjusted Detuning for Ground-Energy Leakage Blockade (ADGLB) method, which leverages an analytical approach to design a laser-detuning sweep schedule based on the spectral gap, without introducing additional Hamiltonian terms, thereby suppressing ground-state population leakage. We validate this approach on $k$-PXP chain-type arrangements, which represent relatively hard instances among problems of comparable size~\cite{Schiffer_PRR_2024}. While direct investigation of the spectral gap is feasible only for small instances, we further show that the optimized schedule can be directly applied to larger instances with similar hardness, as well as to harder instances through modest adjustments of the schedule.

The remainder of this paper is organized as follows. In Sec.~\ref{Theory}, we introduce the theoretical background of adiabatic quantum computing with Rydberg Hamiltonians for MIS preparation along with the ADGLB-based method for scheduling laser detuning. In Sec.~\ref{Results}, we present experimental validations of the ADGLB method, and in Sec.~\ref{Discussion}, we discuss its applicability to more complex problem instances. The conclusions are given in Sec.~\ref{Conclusion}.

\section{Theoretical consideration} \label{Theory}
\subsection{AQC for Rydberg MIS} \label{AQC_Rydberg_MIS} \noindent
Figure~\ref{Figure1}(a) describes an array of $N=7$ Rydberg atoms. Each atom spans a two-level system of the atomic ground state $\ket{g}$ and the Rydberg state $\ket{r}$. When the distance $a$ between adjacent atoms is smaller than $R_b$ (blockade radius), Rydberg blockade occurs, i.e. only one atom can be excited to the Rydberg state $\ket{r}$ under the illumination of a laser. The Hamiltonian $H_{\text{Ry}}$ of such Rydberg atom systems is given by
\begin{equation}
H_{\text{Ry}}(t)= \sum_{v \in V} \left[ \frac{\Omega(t)}{2} \sigma_x^{(v)} + \frac{\delta(t)}{2} \sigma_z^{(v)} \right] + U \sum_{(u, v) \in E} n_u n_v,
\label{H_Ry}
\end{equation}
where $\Omega$ is the Rabi frequency of the laser coupling between $\ket{g}$ and $\ket{r}$, $\delta$ is the laser detuning, $U$ is the van der Waals interaction between two Rydberg atoms of $U=C_6/a^6$ with coefficient $C_6$ and the interatomic distance $a$, and $n_v \equiv \ket{r}_v \bra{r}_v$, $\sigma_x^{(v)}\equiv \ket{r}_v \bra{g}_v + \ket{g}_v \bra{r}_v$ and $\sigma_z^{(v)} \equiv \ket{g}_v \bra{g}_v - \ket{r}_v \bra{r}_v (= 1 - 2 n_v)$ are the excitation number and Pauli-$x, z$ matrices of the node $v$. The Rydberg blockade condition is satisfied when $\Omega  \ll  U$ so that the blockade radius gives that $R_b=(C_6/\Omega)^{1/6}$. The MIS state is given by the lowest-energy eigenstate (ground state) of the Rydberg Hamiltonian $H_{\text{Ry}}$ with $\Omega=0$ and $0< \delta < U$. 

The standard adiabatic evolution schedule for the Rydberg MIS is illustrated in Fig.~\ref{Figure1}~(b). Starting from the initial state where all atoms are in $\ket{g}$ (the ground state of $H_{\text{Ry}}$ with $\Omega=0$ and $\delta=\delta_i <0$), the standard adiabatic schedule consists of the following three stages:

\begin{enumerate}
\item[(i)] The Rabi frequency is ramped up from zero to $\Omega_0$ during the time interval $0\leq t<t_r$.
\item[(ii)] While the Rabi frequency is maintained at $\Omega_0$, the detuning $\delta$ is swept from $\delta_i<0$ to $0<\delta_f<U$, during $t_r \leq t< T-t_r$
\item[(iii)] The Rabi frequency is ramped down from $\Omega_0$ to zero during $T-t_r \leq t \leq T$.
\end{enumerate}
If the evolution under the above schedule is sufficiently adiabatic, the system is always populated in the ground state and reaches the MIS state.

We study the eigenenergy structure and population change for the $N=7$ Rydberg atom array under the standard adiabatic evolution schedule, as in Fig.~\ref{Figure1}~(c-d). The ground-state energy $E_0$, the first excited-state energy $E_1$, and spectral gap $\Delta E_{01} \left( \equiv E_1 - E_0 \right)$ are depicted in Fig.~\ref{Figure1} (c). In the highlighted region in Fig.~\ref{Figure1}(c), the energy gap $\Delta E_{01}$ becomes small compared to the other regions and is minimized at $t=t_\text{min}$ with $\delta_{\text{min}}=\delta(t_{\text{min}})$. In this region, the population $P_{E_0}$ in the ground state starts to leak to higher eigenenergy states, as in Fig.~\ref{Figure1}(d). Then the amount of leakage is $\varepsilon_{P_{E_0}}$.

\begin{figure}[thb!]
    \centering
\includegraphics[width=0.5\textwidth]{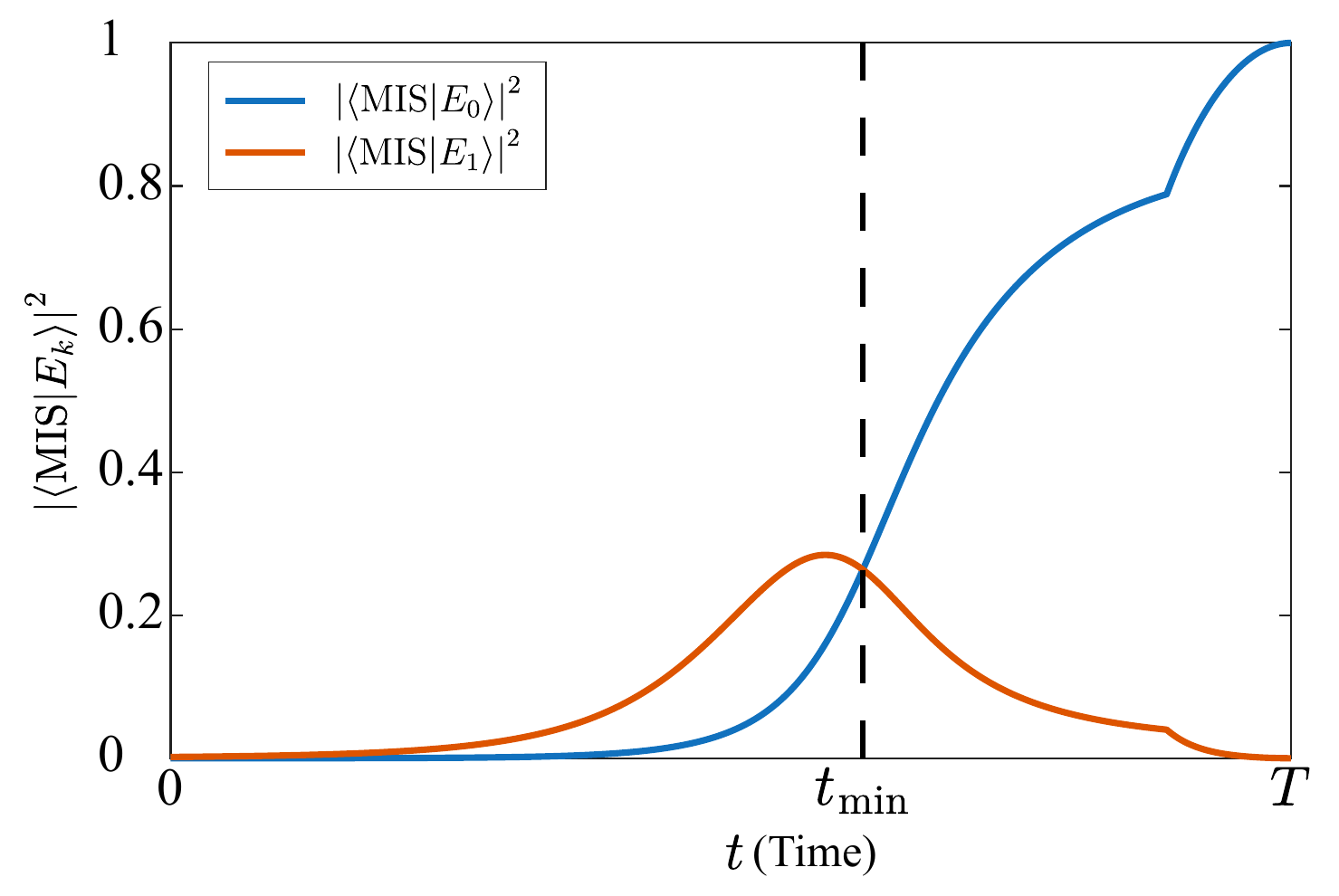}
    \caption{MIS-state overlaps $\left | \bra{\rm MIS}E_k\rangle \right |^2$ with the ground state (blue line) and the first excited state (red line) for the $N=7$ Rydberg atom array in Fig.~\ref{Figure1}(a) along the standard AQC schedule.
    }
\label{Figure2}
\end{figure}

The dynamics of the MIS-state overlap $\left | \bra{\text{MIS}}E_k\rangle \right |^2$ with the ground state, and the first excited state is depicted in Fig.~\ref{Figure2}. In this example, the MIS state is given by $\ket{\text{MIS}} = \ket{rggrggr}$. Although the MIS state perfectly overlaps the ground state at the end of the schedule, mixing between the MIS state and the first excited state occurs near $t=t_{\rm min}$. Consequently, population leakage $\varepsilon_{P_{E_0}}$ from the ground state directly results in infidelity in MIS preparation under the linear detuning schedule $\delta(t)$, indicating that a careful modification of $\delta(t)$ near $t_{\text{min}}$ is required.

For arrangements of the same size, more complex connectivity between Rydberg atoms leads to more complex eigenspectrum structures and a reduced minimum spectral gap $g_{\rm min} \equiv \min_{t \in [0, T]} \Delta E_{01}(t)$, thereby enhancing population leakage from the ground state. Moreover, as the system size increases, the minimum spectral gap $g_{\rm min}$ decreases superexponentially~\cite{Jansen_JMPhys_2007}. The performance of the AQC for a given arrangement can be characterized by the hardness parameter ${\rm HP}=R_{|\text{MIS}|-1}/(|\text{MIS}|R_{|\text{MIS}|})$, where $R_k$ denotes the number of independent sets of size $k$. As a result, the required adiabatic duration $T$ increases for more complex or harder instances, competing with the finite coherence time of the quantum hardware. In the following subsection, we introduce a method to enhance Rydberg AQC performance by exploiting the structural features of the eigenspectrum. We further experimentally demonstrate that a schedule optimized for small instances can be extended to larger arrangements.

\begin{figure*}[thb!]
    \centering
\includegraphics[width=\textwidth]{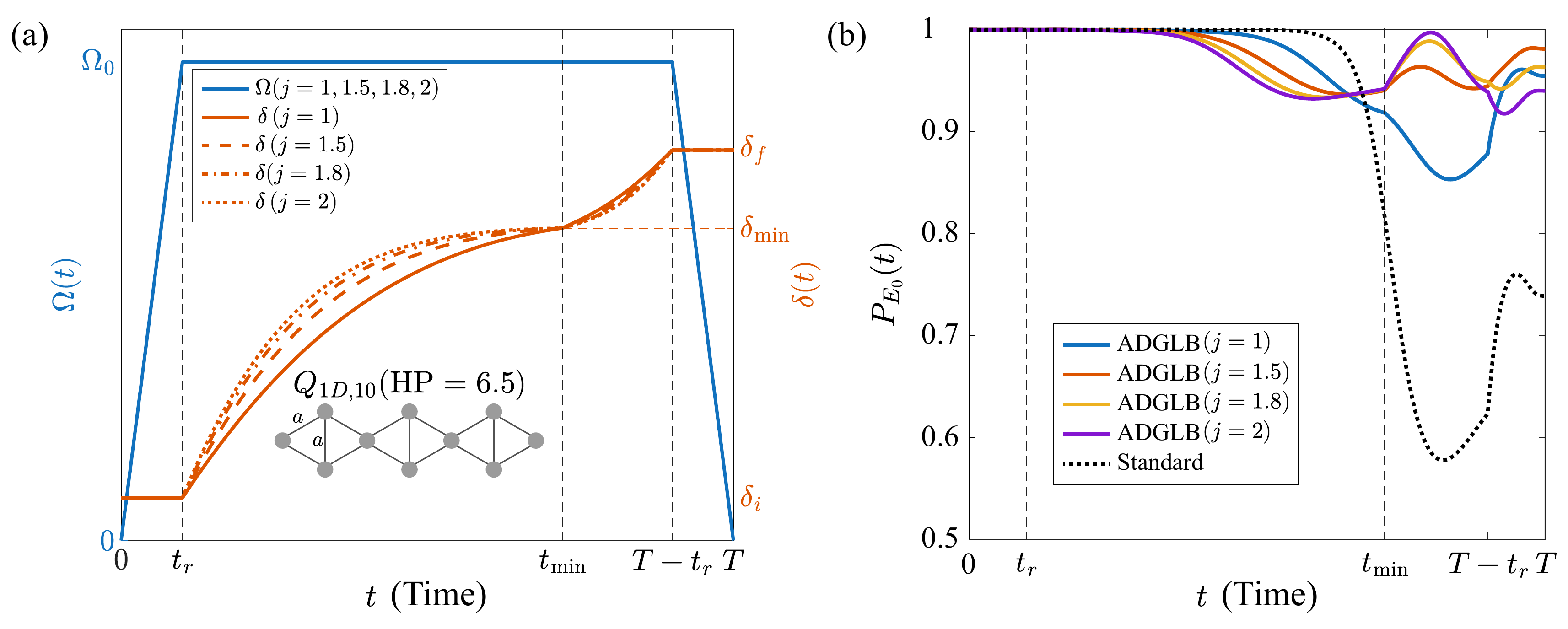}
    \caption{(a) The ADGLB schedule constructed using Eqs.~\eqref{ADGLB_delta_design_eq_zeta_jth_order} and \eqref{ADGLB_delta_design_eq} with $j=1, 1.5, 1.8,$ and $2$, for a quasi-one-dimensional chain of $N=10$ Rydberg atoms, $Q_{1D,10}$ (${\rm HP}=6.5$). (b) Numerical results for the ground-state population dynamics $P_{E_0}(t)$ under the standard schedule (black dashed line) and the ADGLB schedules with $j=1$ (blue solid line), $1.5$ (red solid line), $1.8$ (yellow solid line), and $2$ (purple solid line).}
\label{Figure3}
\end{figure*}

\subsection{Adjusted Detuning for Ground-energy Leakage Blockade (ADGLB)} \label{ADGLB_Pulse_Design} \noindent
The dominant contribution to infidelity in MIS preparation arises from population leakage from the ground state $\ket{E_0}$ to the first excited state $\ket{E_1}$. We therefore consider an effective two-level adiabatic Hamiltonian $H_{\text{ad}}$ defined within the reduced subspace $ \left\{\ket{E_0}, \ket{E_1} \right\}$:

\begin{equation}
H_{\rm ad} = \frac{ \bra{E_1} \frac{dH_{\rm{Ry}}}{dt} \ket{E_0} }{ \Delta E_{01} } s_y - \frac{ \Delta E_{01} }{2} s_z, 
\label{H_new_eq}
\end{equation}
where $s_y \equiv i \ket{E_1} \bra{E_0} - i \ket{E_0} \bra{E_1}$ and $s_z \equiv \ket{E_0} \bra{E_0} - \ket{E_1} \bra{E_1}$ are the Pauli-$y$ and Pauli-$z$ operators in the basis $\{ \ket{E_0}, \ket{E_1} \}$, respectively (See Appendix for details). The time evolution of the amplitudes $\ket{E_0}$ and $\ket{E_1}$ follows from the Schrödinger equation:
\begin{equation} \label{P_E0_TE_Eq}
i \begin{pmatrix}
\dot{c}_0 \\
\dot{c}_1
\end{pmatrix}=H_{\rm ad} \cdot 
\begin{pmatrix}
c_0 \\
c_1
\end{pmatrix},
\end{equation}
where $c_j$ denotes the probability amplitude of $\ket{E_j}$-state ($j=0, 1$), with $P_{E_j}=\lvert c_j \rvert^2$, and $\dot{c}_j\equiv \left(dc_j/dt \right)$ its time-derivative.

In the adiabatic Hamiltonian of Eq.~\eqref{H_new_eq}, the off-diagonal term represents the coupling between $\ket{E_0}$ and $\ket{E_1}$, which induces population leakage from the ground state. Therefore, the coupling term should remain sufficiently small compared to the energy splitting during adiabatic evolution to suppress nonadiabatic transitions. For a fixed evolution time $T$, the spectral gap $\Delta E_{01}$ varies with time and reaches its minimum $g_{\rm min}$. To maintain adiabaticity under this constraint, the rate of change $dH_{\rm Ry}/dt$ should be reduced in the vicinity of $g_{\rm min}$, while it can be increased away from this region compared to the standard schedule. Motivated by this principle, we introduce Adjusted Detuning for Ground-energy Leakage Blockade (ADGLB), a method that modifies the laser-detuning schedule $\delta(t)$ to suppress ground-state leakage. The detailed procedure is described below.

The design of the modified detuning schedule $\delta(t)$ is inspired by the theorem that population leakage $\varepsilon_{P_{E_0}}$ is bounded by the evolution time and inverse power of the spectral gap, $(\Delta E_{01})^{-j}$~\cite{Jansen_JMPhys_2007}. In ADGLB, the Rabi frequency schedule $\Omega(t)$ is kept identical to the standard schedule, while only the detuning $\delta(t)$ is modified within the time interval where the Rabi frequency reaches its maximum value $\Omega_0$ $(t_r<t<T-t_r)$, because the spectral gap $\Delta E_{01}$ is proportional to $\Omega_0$. The detuning is designed such that its time derivative satisfies $d\delta /dt \propto (\Delta E_{01}(t))^j$. The interval is split into two segments at $t_{\rm min}$, where $\Delta E_{01}(t)$ reaches its minimum value $g_{\rm min}$ (at $\delta_{\rm min}$). The detuning path $\delta(t)$ is then constructed using a normalized interpolation function defined as
\begin{equation}
\label{ADGLB_delta_design_eq_zeta_jth_order}
\zeta_j(t, t_0, t_1) \equiv \frac{\int_{t_0}^{t} dt' \cdot [\Delta E_{01}(t')]^j}{\int_{t_0}^{t_1} dt' \cdot [\Delta E_{01}(t')]^j} \quad (t_0 \le t \le t_1,\, j>0). 
\end{equation}
Therefore, the complete detuning schedule is given by
\begin{equation}
\label{ADGLB_delta_design_eq}
\delta(t)=
\begin{cases}
\delta_i \quad \quad \quad \quad \quad \quad \quad \quad \quad 0 \le t \le t_{r}, \\
\delta_i + \left( \delta_{\rm min}-\delta_i \right) \cdot \zeta_j(t, t_{r}, t_{\rm min}) \\
\quad \quad \quad \quad \quad \quad \quad \quad \quad \quad t_{r} \le t \le t_{\rm min}, \\
\delta_{\rm min} + \left( \delta_f-\delta_{\rm min} \right) \cdot \zeta_j(t, t_{\rm min}, T-t_{r}) \\
\quad \quad \quad \quad \quad \quad \quad \quad \quad \quad t_{\rm min} \le t \le T-t_{r}, \\
\delta_f \quad \quad \quad \quad \quad \quad \quad \quad \quad T-t_{r} \le t \le T.
\end{cases}
\end{equation}

We tested ADGLB on a quasi-one-dimensional chain of Rydberg atoms, referred to as the $k$-PXP geometry~\cite{Schiffer_PRR_2024}, whose configuration is shown in Fig.~\ref{Figure3}(a). This geometry exhibits a large number of independent sets of size $|\text{MIS}|-1$, $R_{|\text{MIS}|-1}$, relative to the number of maximum independent sets $R_{|\text{MIS}|}$, resulting in a comparatively large hardness parameter ${\rm HP}$ among arrangements of similar size. Moreover, the triangular geometry provides a clear separation between nearest- and next-nearest-neighbor interactions, which are well described by Eq.~\eqref{H_Ry}. As an example, we determine the modified detuning schedule $\delta(t)$ for a $N=10$ quasi-one-dimensional Rydberg chain, $Q_{1D,10}$, using Eqs.~\eqref{ADGLB_delta_design_eq_zeta_jth_order} and \eqref{ADGLB_delta_design_eq}. The resulting ADGLB schedules for several values of $j$ in the range $1 \leq j \leq 2$ are shown in Fig.~\ref{Figure3}(a).

We compared the ground-state population $P_{E_0}$ under the standard and ADGLB schedules through numerical calculations of the Schrödinger equation with the Hamiltonian in Eq.~\eqref{H_Ry}, as shown in Fig.~\ref{Figure3}(b). While the final ground-state population for the standard schedule was $P_{E_0}=0.739$, the ADGLB schedules yielded $ P_{E_0}> 0.94$. Specifically, for $j=1, 1.5, 1.8,$ and $2$, the corresponding values were $0.955, 0.981, 0.963$, and $0.940$, respectively. Here, the parameter $j$ controls the strength of the schedule modification near the minimum spectral gap. An intermediate value around $j=1.5$ provides the highest performance, reflecting a balance between sufficient slowdown near $g_{\text{min}}$ and avoiding excessive distortion of the overall evolution.

\begin{figure*}[thb!]
    \centering
\includegraphics[width=\textwidth]{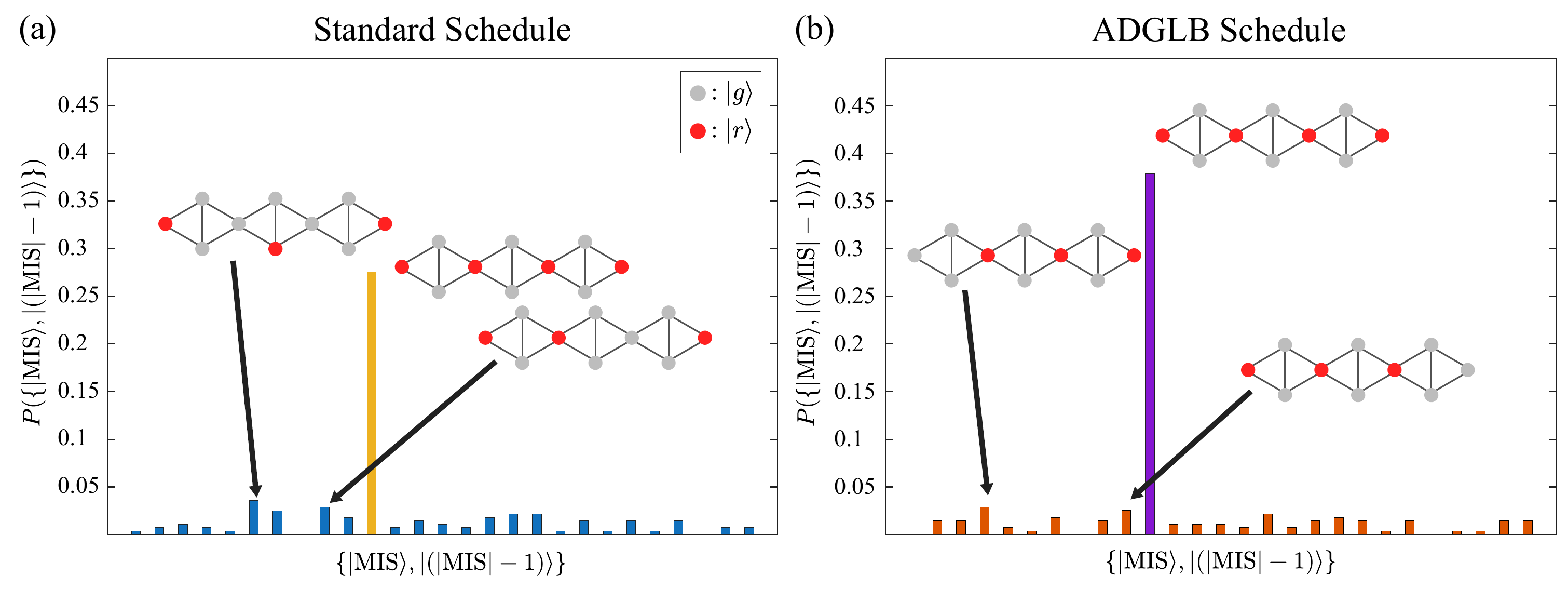}
    \caption{Experimental MIS preparation for the quasi-one-dimensional chain $Q_{1D,10}$. (a) Probability histogram of independent sets of size $|{\rm MIS}|$ (yellow bar) and $|{\rm MIS}|-1$ (blue bars) under the standard schedule (279 valid shots). (b) Probability histogram of independent sets of size $|{\rm MIS}|$ (yellow bar) and $|{\rm MIS}|-1$ (blue bars) under the ADGLB schedule with $j=1.8$ (277 valid shots).
}
\label{Figure4}
\end{figure*}

\section{Results} \label{Results} \noindent
To benchmark ADGLB, we performed experiments on MIS instances using QuEra’s Aquila neutral-atom processor~\cite{Wurtz_Aquila_Arxiv_2023}, accessed via the Amazon Braket service. The laser pulse excites $^{87}$Rb atoms from the atomic ground state $\ket{g}=\ket{5S_{1/2}}$ to the Rydberg state $\ket{r}=\ket{70 S_{1/2}}$, characterized by a van der Waals interaction coefficient $C_6=2\pi\times 863~{\rm GHz \cdot \mu m^6}$. The pulse has a peak Rabi frequency $\Omega_0=2\pi \times 1.0~{\rm MHz}$ and sweeps the detuning from $\delta_i=2\pi \times (-2.5)~{\rm MHz}$ to $\delta_f=2\pi \times (+2.5)~{\rm MHz}$. The nearest-neighbor spacing was $a=8~\mu$m, yielding an interaction strength $U=2\pi \times 3.29~{\rm MHz}$. Under the condition $\delta_f < U$, the MIS configuration corresponds to the ground state of the Rydberg Hamiltonian in Eq.~\eqref{H_Ry}. The ADGLB detuning schedule was implemented with a power exponent $j=1.8$ using Eqs.~\eqref{ADGLB_delta_design_eq_zeta_jth_order} and \eqref{ADGLB_delta_design_eq}. For all experiments, the total adiabatic evolution time was $T=5~\mu$s with a ramp time $t_r=0.5~\mu$s. For the arrangement $Q_{1D,10}$, the minimum-gap parameters are $(t_{\rm min}, \delta_{\rm min}, g_{\rm min})=(3.60~\mu{\rm s}, 2\pi \times 1.38~{\rm MHz}, 2\pi \times 0.29~{\rm MHz})$.

\subsection{Experimental enhancement of MIS preparation in a quasi-one-dimensional chain} \label{Exp_Eval_Q1D} \noindent
We now compare the experimental MIS preparation probabilities under the standard and ADGLB schedules. Figure~\ref{Figure4} shows the experimental probability histograms of independent sets of size $|\text{MIS}|$ and $|\text{MIS}|-1$ for the quasi-one-dimensional chain $Q_{1D,10}$ (${\rm HP}=6.5$), obtained under the standard schedule and the ADGLB schedule with $j=1.8$. The probability of MIS preparation, $P_{\rm MIS}$, increases from 28\% under the standard schedule to 38\% under the ADGLB schedule. Each experiment was repeated 300 times, and the number of valid shots corresponding to defect-free atomic arrays was 279 and 277 for the standard and ADGLB schedules, respectively.

The discrepancy between the numerical results and the experimental observations can be primarily attributed to experimental imperfections of the hardware. These include state-preparation-and-measurement (SPAM) errors of $P(g \vert r)=10.0\%$ and $P(r \vert g)=5.0\%$, a detuning error of $(2\pi)\times 160$~kHz, and a coherence time of $8~\mu$s (measured via Rabi oscillations), limited by residual phase noise of the Rydberg excitation lasers~\cite{Leseleuc_PRA_2018}. Due to these imperfections, the differences in $P({\rm MIS})$ among the ADGLB schedules with different $j$ are not experimentally resolved; however, they remain clearly higher than the probability obtained under the standard schedule.

\begin{figure*}[thb!]
    \centering
\includegraphics[width=\textwidth]{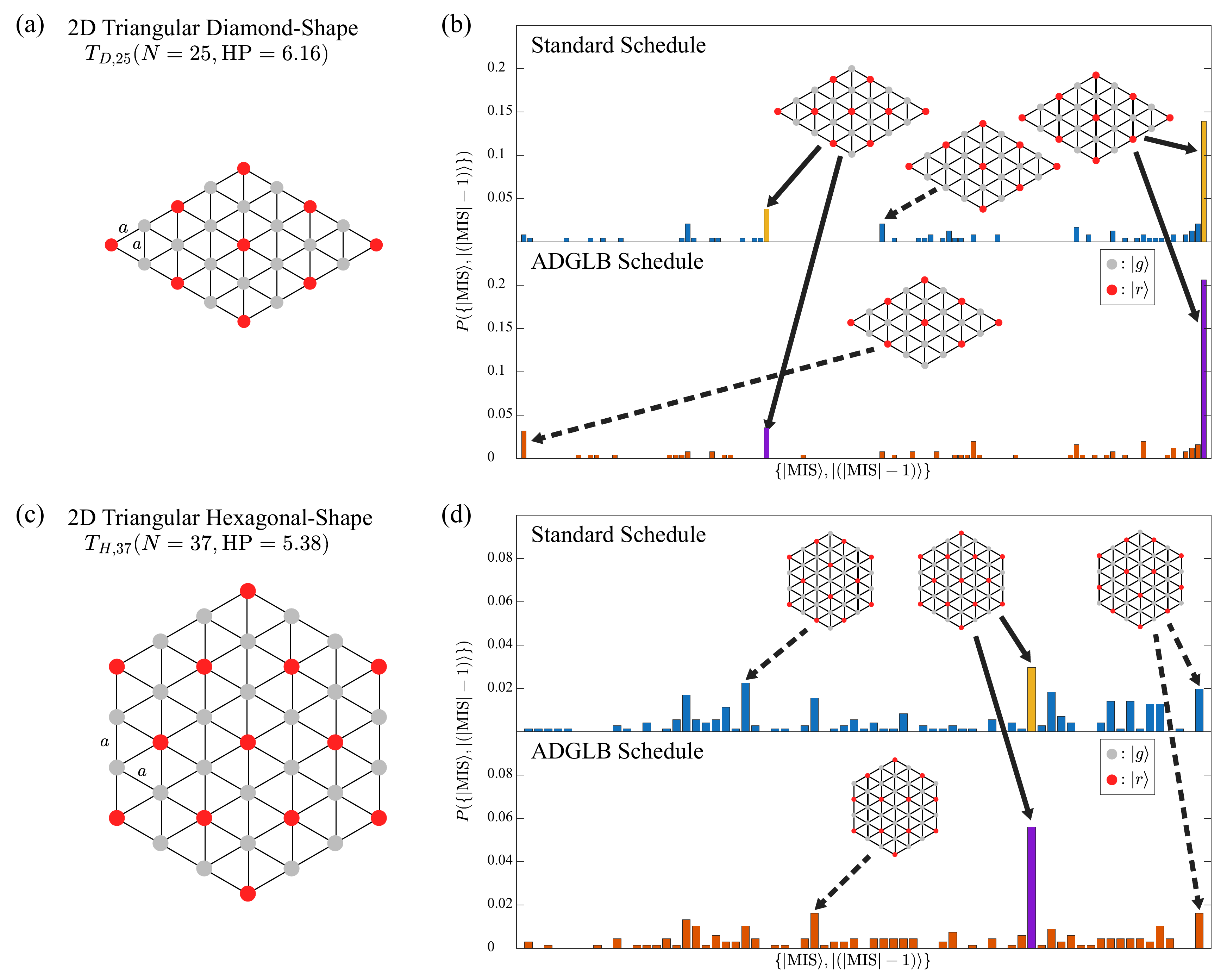}
    \caption{Experimental MIS preparation in two-dimensional triangular lattices. (a) A two-dimensional diamond-shaped lattice $T_{D,25}$. (b) Probability histograms of independent sets of size $|{\rm MIS}|$ and $|{\rm MIS}|-1$ for $T_{D,25}$ under the standard (237 valid shots) and ADGLB (252 valid shots) schedules. (c) A two-dimensional hexagonal-shaped lattice $T_{H,37}$. (d) Probability histograms of independent sets of size $|{\rm MIS}|$ and $|{\rm MIS}|-1$ for $T_{D,25}$ under the standard (707 valid shots) and ADGLB (678 valid shots) schedules. In (b) and (d), yellow and purple bars denote independent sets of size $|{\rm MIS}|$, while blue and red bars denote independent sets of size $|{\rm MIS}|-1$.
}
\label{Figure5}
\end{figure*} 

\subsection{Experimental enhancement of MIS preparation in two-dimensional triangular lattices} \label{Exp_Eval_T2D} \noindent
We demonstrated that the optimal schedule obtained via ADGLB for a quasi-one-dimensional chain can be directly adapted to larger arrangements with similar geometric structures. Figures~\ref{Figure5}(a) and (c) present diamond-shaped and hexagonal-shaped triangular lattice instances, $T_{D,25}$ ($N=25$, ${\rm HP}=6.16$, $\lvert {\rm MIS} \rvert =9$) and $T_{H,37}$ ($N=37$, ${\rm HP}=5.38$, $\lvert {\rm MIS} \rvert =13$), respectively, for which direct diagonalization of the spectral structure becomes computationally demanding. These lattices were constructed by stacking triangular building blocks of the quasi-one-dimensional chain along the perpendicular direction (see Table~\ref{Table_Atom_Coords_of_Tri_2D_Lattices} in Appendix~\ref{Correlation_btw_GS_Leakage_and_E_01} for atomic coordinates). This extension provides a more stringent test of leakage suppression near the minimum spectral gap, owing to the increased connectivity and larger number of $\lvert {\rm MIS} \rvert-1$ configurations.

Both instances have ${\rm HP}$ values similar to those of the quasi-one-dimensional chain $Q_{1D,10}$. The experimental probability histograms for $T_{D,25}$ and $T_{H,37}$ are shown in Fig.~\ref{Figure5}(b) and (d), respectively. The upper panels correspond to the standard schedule, while the lower panels correspond to the ADGLB schedule, using the same schedule as applied in Fig.~\ref{Figure4}. For $T_{D,25}$, the MIS preparation probability $P_{\rm MIS}$ increased from 17.7\% under the standard schedule to 24.2\% under the ADGLB schedule (yellow and purple bars in Fig.~\ref{Figure5}(b), respectively). For $T_{H,37}$, $P_{\rm MIS}$ increased from 3.0\% to 5.6\%. The number of valid shots was 252 (standard: 237) for $T_{D,25}$ and 678 (standard: 707) for $T_{H,37}$.

Overall, these results suggest that schedules optimized using ADGLB for smaller instances can be transferred to larger systems with comparable hardness characteristics. Despite the increased system size and connectivity, applying the same schedule enhances MIS preparation without requiring additional experimental resources or a longer AQC duration. This indicates that the spectral-gap-guided schedule modification effectively reflects the structural features associated with instance hardness rather than merely the system size.

\begin{figure*}[thb!]
    \centering
\includegraphics[width=\textwidth]{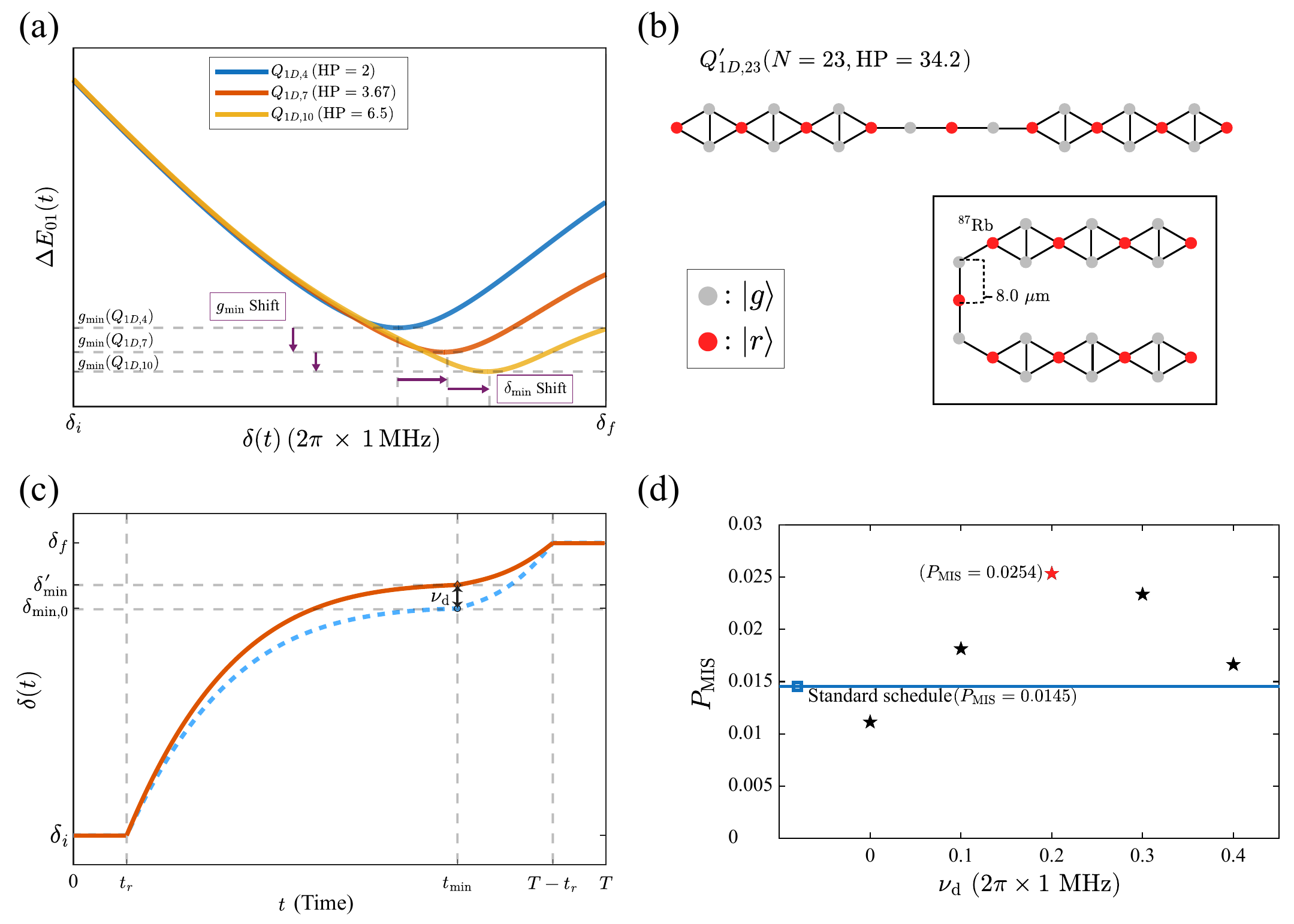}
    \caption{Extension to higher hardness arrangements. 
(a) The spectral gaps  $\Delta E_{01}(t)$ for three quasi-one-dimensional chains with sizes $N=4,\,7,\,10$ ($Q_{1D,4},\, Q_{1D,7}$, and $Q_{1D,10}$ with ${\rm HP}=2,\, 3.67$, and $6.5$, respectively). As $N$ and ${\rm HP}$ increase, the minimum spectral gaps  $g_{\rm min}(Q_{1D, 4}),\, g_{\rm min}(Q_{1D, 7})$, and $g_{\rm min}(Q_{1D, 10})$ decrease, while the corresponding detuning values $\delta_{\rm min}$ increase.
(b) A quasi-one-dimensional chain $Q'_{1D,23}$, consisting of two $Q_{\text{1D},10}$ arrangements connected by a linear atomic chain. The inset depicts the experimental geometry.
(c) The ADGLB schedule for $Q'_{1D,23}$ (red solid line), modified from the ADGLB schedule for $Q_{1D,10}$ (blue dashed line) with a heuristic offset $\nu_{\text{d}}$. 
(d) MIS preparation probability $P_{\rm MIS}$ for $Q'_{1D,23}$ with different heuristic offsets $\nu_{\text{d}}$. The blue solid line denotes $P_{{\rm MIS}}$ under the standard schedule. The red star denotes the maximum MIS preparation probability. The number of valid shots were 894, 899, 882, 907, 899, and 902 for the standard schedule and the modified ADGLB schedules with $\nu_{\text{d}}=0,\,0.1,\,0.2,\,0.3$, and $0.4$, respectively.
}
\label{Figure6}
\end{figure*}

\section{Discussion: Extension to Higher-Hardness Arrangements} \label{Discussion} \noindent
Now we discuss the applicability of the optimized schedule via ADGLB to larger hardness parameter instances. We investigated the typical behavior of the eigenenergy spectra for quasi-one-dimensional arrangements of various sizes. In Fig.~\ref{Figure6}(a), the spectral gaps $\Delta E_{01}(t)$ for arrangements $Q_{\text{1D},4}$, $Q_{\text{1D},7}$, and $Q_{\text{1D},10}$ are shown according to the standard schedule. Not only does the minimum spectral gap $g_{\text{min}}$ decrease for larger hardness parameters, but the corresponding $\delta_{\text{min}}$ also tends to increase.

We show that the schedule optimized via ADGLB for a small-sized, low-hardness arrangement can be applied to a harder arrangement with only a simple modification. As an example, we consider an arrangement $Q'_{\text{1D},23}$ in which it consists of two $Q_{\text{1D},10}$ arrangements connected by a 1D linear atomic chain, as in Fig.~\ref{Figure6}(b). The hardness parameter of the arrangement is $\text{HP}=34.2$, and it is a relatively hard instance with a similar size $N$ compared to the previous two-dimensional triangular lattices. And it is very daunting to estimate $\delta_{\text{min}}$ by directly diagonalizing its Hamiltonian matrix.

Guided by the tendency that $\delta_{\text{min}}$ increases for arrangements with a larger size or HP, we estimate the detuning corresponding to the spectral gap $g_{\text{min}}(Q'_{\text{1D},23})$ by modifying the adiabatic evolution schedule optimized by the ADGLB in the previous section. At $t_{\text{min}}$, we define a modified detuning $\delta_{\text{min}}' =\delta_{\text{min,0}} + \nu_{\text{d}}$, where $\nu_{\text{d}}$ denotes a heuristic offset. The entire modified detuning schedule is given by 
\begin{equation}
\label{ADGLB_delta_design_mod_eq}
\delta(t; \delta'_{\rm min})=
\begin{cases}
\delta_i + \frac{ \delta'_{\rm min} - \delta_i }{ \delta_{\rm min,0} - \delta_i } \cdot \eta_A\left( t - t_r \right) \\
\quad \quad \quad \quad \quad \quad \quad \quad \quad \quad t_{r} \le t \le t_{\rm min}, \\
\delta'_{\rm min} + \frac{ \delta_f - \delta'_{\rm min} }{ \delta_f - \delta_{\rm min,0} } \cdot \eta_B\left( t - t_{\rm min} \right) \\
\quad \quad \quad \quad \quad \quad \quad \quad \quad \quad t_{\rm min} \le t \le T-t_{r}.
\end{cases}
\end{equation}
Here, $\eta_A(s)$ and $\eta_B(s)$ are polynomial approximations of the ADGLB schedule optimized with the $Q_{\text{1D},10}$, expressed as:
\begin{subequations}
\begin{equation}
\eta_A(s) \equiv 4.0047 s -1.5785 s^2 + 0.2826 s^3 - 0.0193 s^4
\end{equation}
\begin{equation}
\eta_B(s) \equiv 0.5509 s - 0.055 s^2 + 1.4402 s^3 - 0.565 s^4,
\end{equation}
\end{subequations}
with a generalized time $s$. The ADGLB and modified detuning schedule $\delta(t)$ are depicted in Fig.~\ref{Figure6}(c) (blue dashed and red solid lines), respectively.

Fig.~\ref{Figure6}~(d) shows the experimental MIS preparation probability $P_{\rm MIS}$ for five different adiabatic evolution schedules defined by Eq.~\eqref{ADGLB_delta_design_mod_eq}, with different values of the heuristic offset $\nu_{\text{d}} = 2\pi\times \left\{0,0.1,0.2,0.3,0.4 \right\}$~MHz. When $\nu_{\text{d}}=0$, we find $P_{\rm MIS}=0.011$, which is smaller than the result obtained using the standard adiabatic evolution schedule ($P_{\rm MIS}=0.0145$, blue line), suggesting that the ADGLB schedule optimized for a small-size arrangement may not be well matched to the eigenenergy structure of larger arrangements. However, as $\nu_{\text{d}}$ increases, $P_{\rm MIS}$ is enhanced and reaches its maximum value of $0.0254$ at $\nu_{\text{d}}=2\pi\times 0.2$~MHz, exceeding the result of the standard adiabatic evolution schedule. The enhancement of $P_{\rm MIS}$ for a finite heuristic offset $\nu_{\text{d}}$ indicates that a slight modification of the detuning near $t_{\text{min}}$ can improve adiabatic evolution near the minimum spectral gap.

\section{Conclusion} \label{Conclusion} \noindent
In summary, we developed a method to enhance MIS preparation in Rydberg-atom adiabatic quantum computations. The proposed ADGLB method leverages the spectral gap of the Rydberg Hamiltonian to modify the laser-detuning schedule, thereby suppressing the population leakage from the ground state without introducing additional Hamiltonian terms. 
We determined an optimized detuning schedule for a quasi-one-dimensional chain of $N=10$ Rydberg atoms, which exhibits a relatively large hardness parameter ${\rm HP}$ compared to arrangements of similar size. The experimental results confirmed that the MIS preparation probability was significantly enhanced relative to the standard schedule. 
Furthermore, we demonstrated that the same ADGLB schedule optimized for the quasi-one-dimensional chain can be directly applied to two-dimensional triangular lattices with $N=25$ and $N=37$, which possess larger system sizes but comparable hardness parameters. In addition, by introducing a heuristic offset to the ADGLB schedule optimized for smaller systems, the method remains effective for instances with higher hardness, including a quasi-one-dimensional chain with $N=23$ and ${\rm HP}=34.2$.

Overall, the spectral-gap-guided schedule modification improved adiabatic performance without requiring repeated use of quantum hardware or complex control terms. By enhancing the quality of the probability distribution generated at the hardware level, this approach may also facilitate classical post-processing and error-mitigation strategies in AQC.

\begin{acknowledgements} \noindent
This research was supported by the National Research Foundation of Korea (NRF) under Grant Nos. RS-2024-00337653 and RS-2025-25464441, funded by the Korea government (MSIT).
\end{acknowledgements}

\section*{Data Availability} \noindent
The data that support the findings of this study are available from the corresponding author upon reasonable request.

\section*{Conflicts of interest} \noindent
The authors declare no conflict of interest.

\bibliography{Refs}

@PREAMBLE{
 "\providecommand{\noopsort}[1]{}" 
 # "\providecommand{\singleletter}[1]{#1}%" 
}

@ARTICLE{JSH_PRR_2023,
  title = {Quantum programming of the satisfiability problem with Rydberg atom graphs},
  author = {Jeong, Seokho and Kim, Minhyuk and Hhan, Minki and Park, JuYoung and Ahn, Jaewook},
  journal = {Phys. Rev. Res.},
  volume = {5},
  issue = {4},
  pages = {043037},
  numpages = {8},
  year = {2023},
  month = {Oct},
  publisher = {American Physical Society},
  doi = {10.1103/PhysRevResearch.5.043037},
  url = {https://link.aps.org/doi/10.1103/PhysRevResearch.5.043037}
}

@ARTICLE{Jansen_JMPhys_2007,
    author = {Jansen, Sabine and Ruskai, Mary-Beth and Seiler, Ruedi},
    title = {Bounds for the adiabatic approximation with applications to quantum computation},
    journal = {J. Math. Phys.},
    volume = {48},
    number = {10},
    pages = {102111},
    year = {2007},
    month = {10},
    issn = {0022-2488},
    doi = {10.1063/1.2798382},
    url = {https://doi.org/10.1063/1.2798382},
}

@ARTICLE{Schiffer_PRR_2024,
  title = {Circumventing superexponential runtimes for hard instances of quantum adiabatic optimization},
  author = {Schiffer, Benjamin F. and Wild, Dominik S. and Maskara, Nishad and Cain, Madelyn and Lukin, Mikhail D. and Samajdar, Rhine},
  journal = {Phys. Rev. Res.},
  volume = {6},
  issue = {1},
  pages = {013271},
  numpages = {14},
  year = {2024},
  month = {Mar},
  publisher = {American Physical Society},
  doi = {10.1103/PhysRevResearch.6.013271},
  url = {https://link.aps.org/doi/10.1103/PhysRevResearch.6.013271}
}

@MISC{Wurtz_Aquila_Arxiv_2023,
      title={Aquila: QuEra's 256-qubit neutral-atom quantum computer}, 
      author={Jonathan Wurtz and Alexei Bylinskii and Boris Braverman and Jesse Amato-Grill and Sergio H. Cantu and Florian Huber and Alexander Lukin and Fangli Liu and Phillip Weinberg and John Long and Sheng-Tao Wang and Nathan Gemelke and Alexander Keesling},
      year={2023},
      eprint={2306.11727},
      archivePrefix={arXiv},
      url={https://arxiv.org/abs/2306.11727}, 
}

@ARTICLE{Leseleuc_PRA_2018,
  title = {Analysis of imperfections in the coherent optical excitation of single atoms to Rydberg states},
  author = {de L\'es\'eleuc, Sylvain and Barredo, Daniel and Lienhard, Vincent and Browaeys, Antoine and Lahaye, Thierry},
  journal = {Phys. Rev. A},
  volume = {97},
  issue = {5},
  pages = {053803},
  pages = {9},
  year = {2018},
  month = {May},
  publisher = {American Physical Society},
  doi = {10.1103/PhysRevA.97.053803},
  url = {https://link.aps.org/doi/10.1103/PhysRevA.97.053803}
}

@ARTICLE{Leseleuc_Science_2019,
author = {de L\'es\'eleuc, Sylvain  and Lienhard, Vincent and Scholl, Pascal and Barredo, Daniel and Weber, Sebastian and Lang, Nicolai and others},
title = {Observation of a symmetry-protected topological phase of interacting bosons with Rydberg atoms},
journal = {Science},
volume = {365},
number = {6455},
pages = {775-780},
year = {2019},
doi = {10.1126/science.aav9105},
URL = {https://www.science.org/doi/abs/10.1126/science.aav9105}}

@ARTICLE{JSH_AQT_2025,
  author = {Jeong, Seokho and Park, Juyoung and Ahn, Jaewook},
  title = {Quantum-Enhanced Simulated Annealing Using Rydberg Atoms},
  journal = {Adv. Quantum Technol.},
  volume = {8},
  number = {12},
  pages = {e2500070},
  year = {2025},
  doi = {https://doi.org/10.1002/qute.202500070},
  url = {https://advanced.onlinelibrary.wiley.com/doi/abs/10.1002/qute.202500070},
  year = {2025}
}

@ARTICLE{Feynman_IJTP_1982,
  author = {Feynman, Richard P},
  title = {Simulating Physics with Computers},
  journal = {Int. J. Theor. Phys.},
  volume = {21},
  number = {6/7},
  pages = {467-488},
  year = {1982},
  doi = {10.1007/BF02650179},
  url = {https://link.springer.com/article/10.1007/BF02650179}
}

@BOOK{Nielsen_Chuang_Book_2010,
  title={Quantum computation and quantum information},
  author={Nielsen, Michael A and Chuang, Isaac L},
  year={2010},
  publisher={Cambridge university press}
}

@ARTICLE{Dowling_PTRSA_2003,
    author = {MacFarlane, A. G. J. and Dowling, Jonathan P. and Milburn, Gerard J.},
    title = {Quantum technology: the second quantum revolution},
    journal = {Philos. Trans. A. Math. Phys. and Eng. Sci.},
    volume = {361},
    number = {1809},
    pages = {1655-1674},
    year = {2003},
    month = {06},
    issn = {1364-503X},
    doi = {10.1098/rsta.2003.1227},
    url = {https://doi.org/10.1098/rsta.2003.1227},
}

@INPROCEEDINGS{Shor_IEEE_1994,
  author={Shor, P.W.},
  booktitle={Proc. 35th Ann. Symp. on Foundations of Computer Science}, 
  title={Algorithms for quantum computation: discrete logarithms and factoring}, 
  year={1994},
  volume={},
  number={},
  pages={124-134},
  doi={10.1109/SFCS.1994.365700}}

@ARTICLE{Grover_PRL_1997,
  title = {Quantum Mechanics Helps in Searching for a Needle in a Haystack},
  author = {Grover, Lov K.},
  journal = {Phys. Rev. Lett.},
  volume = {79},
  issue = {2},
  pages = {325--328},
  numpages = {0},
  year = {1997},
  month = {Jul},
  publisher = {American Physical Society},
  doi = {10.1103/PhysRevLett.79.325},
  url = {https://link.aps.org/doi/10.1103/PhysRevLett.79.325}
}

@ARTICLE{Farhi_Science_2001,
author = {Edward Farhi  and Jeffrey Goldstone  and Sam Gutmann  and Joshua Lapan  and Andrew Lundgren  and Daniel Preda },
title = {A Quantum Adiabatic Evolution Algorithm Applied to Random Instances of an NP-Complete Problem},
journal = {Science},
volume = {292},
number = {5516},
pages = {472-475},
year = {2001},
doi = {10.1126/science.1057726},
URL = {https://www.science.org/doi/abs/10.1126/science.1057726}}

@ARTICLE{Dickson_PRL_2011,
  title = {Does Adiabatic Quantum Optimization Fail for NP-Complete Problems?},
  author = {Dickson, Neil G. and Amin, M. H. S.},
  journal = {Phys. Rev. Lett.},
  volume = {106},
  issue = {5},
  pages = {050502},
  numpages = {4},
  year = {2011},
  month = {Feb},
  publisher = {American Physical Society},
  doi = {10.1103/PhysRevLett.106.050502},
  url = {https://link.aps.org/doi/10.1103/PhysRevLett.106.050502}
}

@ARTICLE{Monroe_RMP_2021,
  title = {Programmable quantum simulations of spin systems with trapped ions},
  author = {Monroe, C. and Campbell, W. C. and Duan, L.-M. and Gong, Z.-X. and Gorshkov, A. V. and Hess, P. W. and others},
  journal = {Rev. Mod. Phys.},
  volume = {93},
  issue = {2},
  pages = {025001},
  numpages = {57},
  year = {2021},
  month = {Apr},
  publisher = {American Physical Society},
  doi = {10.1103/RevModPhys.93.025001},
  url = {https://link.aps.org/doi/10.1103/RevModPhys.93.025001}
}

@ARTICLE{Arute_Nature_2019,
  title={Quantum supremacy using a programmable superconducting processor},
  author={Arute, Frank and Arya, Kunal and Babbush, Ryan and Bacon, Dave and Bardin, Joseph C and Barends, Rami and others},
  journal={Nature},
  volume={574},
  number={7779},
  pages={505--510},
  year={2019},
  publisher={Nature Publishing Group UK London},
  doi={10.1038/s41586-019-1666-5},
  url={https://www.nature.com/articles/s41586-019-1666-5}
}

@ARTICLE{Ebadi_Nature_2021,
  title={Quantum phases of matter on a 256-atom programmable quantum simulator},
  author={Ebadi, Sepehr and Wang, Tout T and Levine, Harry and Keesling, Alexander and Semeghini, Giulia and Omran, Ahmed and others},
  journal={Nature},
  volume={595},
  number={7866},
  pages={227--232},
  year={2021},
  publisher={Nature Publishing Group UK London},
  doi={10.1038/s41586-021-03582-4},
  url={https://www.nature.com/articles/s41586-021-03582-4}
}

@ARTICLE{Saffman_RMP_2010,
  title = {Quantum information with Rydberg atoms},
  author = {Saffman, M. and Walker, T. G. and M\o{}lmer, K.},
  journal = {Rev. Mod. Phys.},
  volume = {82},
  issue = {3},
  pages = {2313--2363},
  numpages = {0},
  year = {2010},
  month = {Aug},
  publisher = {American Physical Society},
  doi = {10.1103/RevModPhys.82.2313},
  url = {https://link.aps.org/doi/10.1103/RevModPhys.82.2313}
}

@BOOK{Gallagher_Book_1994, 
place={Cambridge}, 
series={Cambridge Monographs on Atomic, Molecular and Chemical Physics}, 
title={Rydberg Atoms}, 
publisher={Cambridge University Press}, author={Gallagher, Thomas F.}, 
year={1994}, 
collection={Cambridge Monographs on Atomic, Molecular and Chemical Physics}}

@ARTICLE{Endres_Science_2016,
author = {Manuel Endres  and Hannes Bernien  and Alexander Keesling  and Harry Levine  and Eric R. Anschuetz  and Alexandre Krajenbrink  and others},
title = {Atom-by-atom assembly of defect-free one-dimensional cold atom arrays},
journal = {Science},
volume = {354},
number = {6315},
pages = {1024-1027},
year = {2016},
doi = {10.1126/science.aah3752},
url = {https://www.science.org/doi/abs/10.1126/science.aah3752}}

@ARTICLE{Adams_JPhysB_2019,
doi = {10.1088/1361-6455/ab52ef},
url = {https://doi.org/10.1088/1361-6455/ab52ef},
year = {2019},
month = {dec},
publisher = {IOP Publishing},
volume = {53},
number = {1},
pages = {012002},
author = {Adams, C S and Pritchard, J D and Shaffer, J P},
title = {Rydberg atom quantum technologies},
journal = {J. Phys. B: At. Mol. Opt. Phys.}
}

@ARTICLE{Saffman_JPhysB_2016,
doi = {10.1088/0953-4075/49/20/202001},
url = {https://doi.org/10.1088/0953-4075/49/20/202001},
year = {2016},
month = {oct},
publisher = {IOP Publishing},
volume = {49},
number = {20},
pages = {202001},
author = {Saffman, M},
title = {Quantum computing with atomic qubits and Rydberg interactions: progress and challenges},
journal = {J. Phys. B: At. Mol. Opt. Phys.}
}

@ARTICLE{Barik_FQST_2024,
  author={Barik, Shovan Kanti  and Thakur, Aishwarya  and Jindal, Yashica  and S, Silpa B.  and Roy, Sanjukta },
  title={Quantum technologies with Rydberg atoms},
  journal={Front. Quantum Sci. Technol.},
  volume={3},
  pages = {1426216},
  year={2024},  
  url={https://www.frontiersin.org/journals/quantum-science-and-technology/articles/10.3389/frqst.2024.1426216},
  doi={10.3389/frqst.2024.1426216},
  issn={2813-2181}}

@ARTICLE{Morgado_AVSQuantSci_2021,
    author = {Morgado, M. and Whitlock, S.},
    title = {Quantum simulation and computing with Rydberg-interacting qubits},
    journal = {AVS Quantum Sci.},
    volume = {3},
    number = {2},
    pages = {023501},
    year = {2021},
    month = {05},
    issn = {2639-0213},
    doi = {10.1116/5.0036562},
    url = {https://doi.org/10.1116/5.0036562},
    publisher={AIP Publishing}
}

@MISC{Pichler_Arxiv_2018,
      title={Quantum Optimization for Maximum Independent Set Using Rydberg Atom Arrays}, 
      author={Hannes Pichler and Sheng-Tao Wang and Leo Zhou and Soonwon Choi and Mikhail D. Lukin},
      year={2018},
      eprint={1808.10816},
      archivePrefix={arXiv},
      url={https://arxiv.org/abs/1808.10816}, 
}

@ARTICLE{Pichler_PRL_2025,
  title = {Error-Corrected Fermionic Quantum Processors with Neutral Atoms},
  author = {Ott, Robert and Gonz\'alez-Cuadra, Daniel and Zache, Torsten V. and Zoller, Peter and Kaufman, Adam M. and Pichler, Hannes},
  journal = {Phys. Rev. Lett.},
  volume = {135},
  issue = {9},
  pages = {090601},
  numpages = {7},
  year = {2025},
  month = {Aug},
  publisher = {American Physical Society},
  doi = {10.1103/zkpl-hh28},
  url = {https://link.aps.org/doi/10.1103/zkpl-hh28}
}

@ARTICLE{MHKim_NPhys_2022,
  title={Rydberg quantum wires for maximum independent set problems},
  author={Kim, Minhyuk and Kim, Kangheun and Hwang, Jaeyong and Moon, Eun-Gook and Ahn, Jaewook},
  journal={Nature Physics},
  volume={18},
  number={7},
  pages={755--759},
  year={2022},
  publisher={Nature Publishing Group UK London},
  doi = {10.1038/s41567-022-01629-5},
  url = {https://www.nature.com/articles/s41567-022-01629-5}
}

@ARTICLE{Dalyac_PRA_2023,
  title = {Exploring the impact of graph locality for the resolution of the maximum- independent-set problem with neutral atom devices},
  author = {Dalyac, Constantin and Henry, Louis-Paul and Kim, Minhyuk and Ahn, Jaewook and Henriet, Lo\"{\i}c},
  journal = {Phys. Rev. A},
  volume = {108},
  issue = {5},
  pages = {052423},
  numpages = {7},
  year = {2023},
  month = {Nov},
  publisher = {American Physical Society},
  doi = {10.1103/PhysRevA.108.052423},
  url = {https://link.aps.org/doi/10.1103/PhysRevA.108.052423}
}

@ARTICLE{AByun_PRXQ_2022,
  title = {Finding the Maximum Independent Sets of Platonic Graphs Using Rydberg Atoms},
  author = {Byun, Andrew and Kim, Minhyuk and Ahn, Jaewook},
  journal = {PRX Quantum},
  volume = {3},
  issue = {3},
  pages = {030305},
  numpages = {10},
  year = {2022},
  month = {Jul},
  publisher = {American Physical Society},
  doi = {10.1103/PRXQuantum.3.030305},
  url = {https://link.aps.org/doi/10.1103/PRXQuantum.3.030305}
}

@ARTICLE{KKH_SD_2024,
  title={Quantum computing dataset of maximum independent set problem on king lattice of over hundred Rydberg atoms},
  author={Kim, Kangheun and Kim, Minhyuk and Park, Juyoung and Byun, Andrew and Ahn, Jaewook},
  journal={Sci. Data},
  volume={11},
  number={1},
  pages={111},
  year={2024},
  publisher={Nature Publishing Group UK London},
  doi = {10.1038/s41597-024-02926-9},
  url = {https://www.nature.com/articles/s41597-024-02926-9}
}

@ARTICLE{PJY_PRR_2024,
  title = {Rydberg-atom experiment for the integer factorization problem},
  author = {Park, Juyoung and Jeong, Seokho and Kim, Minhyuk and Kim, Kangheun and Byun, Andrew and Vignoli, Louis and others},
  journal = {Phys. Rev. Res.},
  volume = {6},
  issue = {2},
  pages = {023241},
  numpages = {12},
  year = {2024},
  month = {Jun},
  publisher = {American Physical Society},
  doi = {10.1103/PhysRevResearch.6.023241},
  url = {https://link.aps.org/doi/10.1103/PhysRevResearch.6.023241}
}

@ARTICLE{Ebadi_Science_2022,
author = {Ebadi, Sepehr and Keesling, Alexander and Cain, Madelyn and Wang, Tout T and Levine, Harry and Bluvstein, Dolev and others},
title = {Quantum optimization of maximum independent set using Rydberg atom arrays},
journal = {Science},
volume = {376},
number = {6598},
pages = {1209-1215},
year = {2022},
doi = {10.1126/science.abo6587},
url = {https://www.science.org/doi/abs/10.1126/science.abo6587}}

@ARTICLE{YSong_PRR_2021,
  title = {Quantum simulation of Cayley-tree Ising Hamiltonians with three-dimensional Rydberg atoms},
  author = {Song, Yunheung and Kim, Minhyuk and Hwang, Hansub and Lee, Woojun and Ahn, Jaewook},
  journal = {Phys. Rev. Res.},
  volume = {3},
  issue = {1},
  pages = {013286},
  numpages = {7},
  year = {2021},
  month = {Mar},
  publisher = {American Physical Society},
  doi = {10.1103/PhysRevResearch.3.013286},
  url = {https://link.aps.org/doi/10.1103/PhysRevResearch.3.013286}
}

@ARTICLE{AByun_AQT_2024,
author = {Byun, Andrew and Jung, Junwoo and Kim, Kangheun and Kim, Minhyuk and Jeong, Seokho and Jeong, Heejeong and others},
title = {Rydberg-Atom Graphs for Quadratic Unconstrained Binary Optimization Problems},
journal = {Adv Quantum Technol.},
volume = {7},
number = {8},
pages = {2300398},
doi = {https://doi.org/10.1002/qute.202300398},
url = {https://advanced.onlinelibrary.wiley.com/doi/abs/10.1002/qute.202300398},
year = {2024}
}

@ARTICLE{Oliveira_PRXQ_2025,
  title = {Demonstration of Weighted-Graph Optimization on a Rydberg-Atom Array Using Local Light Shifts},
  author = {de Oliveira, A. G. and Diamond-Hitchcock, E. and Walker, D. M. and Wells-Pestell, M. T. and Pelegr\'{\i}, G. and Picken, C. J. and others},
  journal = {PRX Quantum},
  volume = {6},
  issue = {1},
  pages = {010301},
  numpages = {12},
  year = {2025},
  month = {Jan},
  publisher = {American Physical Society},
  doi = {10.1103/PRXQuantum.6.010301},
  url = {https://link.aps.org/doi/10.1103/PRXQuantum.6.010301}
}

@ARTICLE{Finzgar_PRR_2024,
  title = {Designing quantum annealing schedules using Bayesian optimization},
  author = {Fin\ifmmode \check{z}\else \v{z}\fi{}gar, Jernej Rudi and Schuetz, Martin J. A. and Brubaker, J. Kyle and Nishimori, Hidetoshi and Katzgraber, Helmut G.},
  journal = {Phys. Rev. Res.},
  volume = {6},
  issue = {2},
  pages = {023063},
  numpages = {17},
  year = {2024},
  month = {Apr},
  publisher = {American Physical Society},
  doi = {10.1103/PhysRevResearch.6.023063},
  url = {https://link.aps.org/doi/10.1103/PhysRevResearch.6.023063}
}

@ARTICLE{Perseguers_PRApplied_2025,
  title = {Hardness-dependent quantum adiabatic schedules for the maximum-independent-set problem},
  author = {Perseguers, S\'ebastien},
  journal = {Phys. Rev. Appl.},
  volume = {23},
  issue = {6},
  pages = {064023},
  numpages = {15},
  year = {2025},
  month = {Jun},
  publisher = {American Physical Society},
  doi = {10.1103/PhysRevApplied.23.064023},
  url = {https://link.aps.org/doi/10.1103/PhysRevApplied.23.064023}
}

@MISC{Hsieh_Arxiv_2025,
      title={Less is more: subspace reduction for counterdiabatic driving of Rydberg atom arrays}, 
      author={Wen Ting Hsieh and Dries Sels},
      year={2025},
      eprint={2512.04494},
      archivePrefix={arXiv},
      url={https://arxiv.org/abs/2512.04494}, 
}

@MISC{Zhang_Arxiv_2024,
      title={Analog Counterdiabatic Quantum Computing}, 
      author={Qi Zhang and Narendra N. Hegade and Alejandro Gomez Cadavid and Lucas Lassablière and Jan Trautmann and Sébastien Perseguers and others},
      year={2024},
      eprint={2405.14829},
      archivePrefix={arXiv},
      url={https://arxiv.org/abs/2405.14829}, 
}

@ARTICLE{Braida_Quantum_2025,
  doi = {10.22331/q-2025-07-11-1790},
  url = {https://doi.org/10.22331/q-2025-07-11-1790},
  title = {Unstructured {A}diabatic {Q}uantum {O}ptimization: {O}ptimality with {L}imitations},
  author = {Braida, Arthur and Chakraborty, Shantanav and Chaudhuri, Alapan and Cunningham, Joseph and Menavlikar, Rutvij and Novo, Leonardo and others},
  journal = {{Quantum}},
  issn = {2521-327X},
  publisher = {{Verein zur F{\"{o}}rderung des Open Access Publizierens in den Quantenwissenschaften}},
  volume = {9},
  pages = {1790},
  month = jul,
  year = {2025}
}

@ARTICLE{Mozgunov_PTRSA_2022,
    author = {Mozgunov, Evgeny and Lidar, Daniel A.},
    title = {Quantum adiabatic theorem for unbounded Hamiltonians with a cutoff and its application to superconducting circuits},
    journal = {Philos. Trans. A. Math. Phys. and Eng. Sci.},
    volume = {381},
    number = {2241},
    pages = {20210407},
    year = {2022},
    month = {12},
    issn = {1364-503X},
    doi = {10.1098/rsta.2021.0407},
    url = {https://doi.org/10.1098/rsta.2021.0407},
}

@MISC{VickyChoi_Arxiv_2010,
      title={Adiabatic Quantum Algorithms for the NP-Complete Maximum-Weight Independent Set, Exact Cover and 3SAT Problems}, 
      author={Vicky Choi},
      year={2010},
      eprint={1004.2226},
      archivePrefix={arXiv},
      url={https://arxiv.org/abs/1004.2226}, 
}

@ARTICLE{Dupont_SciAdv_2023,
author = {Dupont, Maxime and Evert, Bram and Hodson, Mark J. and Sundar, Bhuvanesh and Jeffrey, Stephen and Yamaguchi, Yuki and others},
title = {Quantum-enhanced greedy combinatorial optimization solver},
journal = {Sci. Adv.},
volume = {9},
number = {45},
pages = {eadi0487},
year = {2023},
doi = {10.1126/sciadv.adi0487},
URL = {https://www.science.org/doi/abs/10.1126/sciadv.adi0487}}

@ARTICLE{Dupont_PRAppl_2025,
  title = {Benchmarking quantum optimization for the maximum-cut problem on a superconducting quantum computer},
  author = {Dupont, Maxime and Sundar, Bhuvanesh and Evert, Bram and Neira, David E. Bernal and Peng, Zedong and Jeffrey, Stephen and others},
  journal = {Phys. Rev. Appl.},
  volume = {23},
  issue = {1},
  pages = {014045},
  numpages = {38},
  year = {2025},
  month = {Jan},
  publisher = {American Physical Society},
  doi = {10.1103/PhysRevApplied.23.014045},
  url = {https://link.aps.org/doi/10.1103/PhysRevApplied.23.014045}
}

@ARTICLE{Niroula_SD_2022,
  title={Constrained quantum optimization for extractive summarization on a trapped-ion quantum computer},
  author={Niroula, Pradeep and Shaydulin, Ruslan and Yalovetzky, Romina and Minssen, Pierre and Herman, Dylan and Hu, Shaohan and others},
  journal={Sci. Rep.},
  volume={12},
  number={1},
  pages={17171},
  year={2022},
  doi = {https://doi.org/10.1038/s41598-022-20853-w},
  url = {https://www.nature.com/articles/s41598-022-20853-w},
  publisher={Nature Publishing Group UK London}
}

@ARTICLE{Zhu_QST_2023,
doi = {10.1088/2058-9565/ac91ef},
url = {https://doi.org/10.1088/2058-9565/ac91ef},
year = {2022},
month = {nov},
publisher = {IOP Publishing},
volume = {8},
number = {1},
pages = {015007},
author = {Zhu, Yingyue and Zhang, Zewen and Sundar, Bhuvanesh and Green, Alaina M and Huerta Alderete, C and Nguyen, Nhung H and others},
title = {Multi-round QAOA and advanced mixers on a trapped-ion quantum computer},
journal = {Quantum Sci. Technol.}
}

@ARTICLE{Zhang_Science_2023,
author = {Zhang, Xueyue  and Kim, Eunjong and Mark, Daniel K. and Choi, Soonwon and Painter, Oskar},
title = {A superconducting quantum simulator based on a photonic-bandgap metamaterial},
journal = {Science},
volume = {379},
number = {6629},
pages = {278-283},
year = {2023},
doi = {10.1126/science.ade7651},
URL = {https://www.science.org/doi/abs/10.1126/science.ade7651}
}

@article{Fauseweh_NatComms_2024,
  title={Quantum many-body simulations on digital quantum computers: State-of-the-art and future challenges},
  author={Fauseweh, Benedikt},
  journal={Nat. Commun.},
  volume={15},
  number={1},
  pages={2123},
  year={2024},
  doi = {10.1038/s41467-024-46402-9},
  URL = {https://www.nature.com/articles/s41467-024-46402-9},
  publisher={Nature Publishing Group UK London}
}

@article{Satzinger_Science_2021,
author = {Satzinger, K. J. and Liu, Y.-J and Smith, A. and Knapp, C. and Newman, M. and Jones, C. and others},
title = {Realizing topologically ordered states on a quantum processor},
journal = {Science},
volume = {374},
number = {6572},
pages = {1237-1241},
year = {2021},
doi = {10.1126/science.abi8378},
URL = {https://www.science.org/doi/abs/10.1126/science.abi8378}
}

@article{Kaplan_PRL_2020,
  title = {Many-Body Dephasing in a Trapped-Ion Quantum Simulator},
  author = {Kaplan, Harvey B. and Guo, Lingzhen and Tan, Wen Lin and De, Arinjoy and Marquardt, Florian and Pagano, Guido and others},
  journal = {Phys. Rev. Lett.},
  volume = {125},
  issue = {12},
  pages = {120605},
  numpages = {7},
  year = {2020},
  month = {Sep},
  publisher = {American Physical Society},
  doi = {10.1103/PhysRevLett.125.120605},
  url = {https://link.aps.org/doi/10.1103/PhysRevLett.125.120605}
}

@article{Wang_PRA_2012,
  title = {Intrinsic phonon effects on analog quantum simulators with ultracold trapped ions},
  author = {Wang, C.-C. Joseph and Freericks, J. K.},
  journal = {Phys. Rev. A},
  volume = {86},
  issue = {3},
  pages = {032329},
  numpages = {20},
  year = {2012},
  month = {Sep},
  publisher = {American Physical Society},
  doi = {10.1103/PhysRevA.86.032329},
  url = {https://link.aps.org/doi/10.1103/PhysRevA.86.032329}
}

@article{Pagano_QST_2019,
doi = {10.1088/2058-9565/aae0fe},
url = {https://doi.org/10.1088/2058-9565/aae0fe},
year = {2018},
month = {oct},
publisher = {IOP Publishing},
volume = {4},
number = {1},
pages = {014004},
author = {Pagano, G and Hess, P W and Kaplan, H B and Tan, W L and Richerme, P and Becker, P and others},
title = {Cryogenic trapped-ion system for large scale quantum simulation},
journal = {Quantum Sci. Technol.}
}

@article{Acharya_Nature_2025,
  title={Quantum error correction below the surface code threshold},
  author={Acharya, R. and Abanin, Dmitry A. and Aghababaie-Beni, Laleh and Aleiner, Igor and Andersen, Trond I. and Ansmann, Markus and others},
  journal={Nature},
  volume={638},
  number={8052},
  pages={920--926},
  year={2025},
  doi = {10.1038/s41586-024-08449-y},
  url = {https://www.nature.com/articles/s41586-024-08449-y},
  publisher={Nature Publishing Group UK London}
}

@article{Bluvstein_Nature_2026,
  title={A fault-tolerant neutral-atom architecture for universal quantum computation},
  author={Bluvstein, Dolev and Geim, Alexandra A and Li, Sophie H and Evered, Simon J and Bonilla Ataides, J Pablo and Baranes, Gefen and and others},
  journal={Nature},
  volume={649},
  pages={39--46},
  year={2026},
  doi = {10.1038/s41586-025-09848-5},
  url = {https://www.nature.com/articles/s41586-025-09848-5},
  publisher={Nature Publishing Group UK London}
}

@article{RodriguezBlanco_PRA_2024,
  title = {Witnessing entanglement in trapped-ion quantum error correction under realistic noise},
  author = {Rodriguez-Blanco, Andrea and Shahandeh, Farid and Bermudez, Alejandro},
  journal = {Phys. Rev. A},
  volume = {109},
  issue = {5},
  pages = {052417},
  numpages = {29},
  year = {2024},
  month = {May},
  publisher = {American Physical Society},
  doi = {10.1103/PhysRevA.109.052417},
  url = {https://link.aps.org/doi/10.1103/PhysRevA.109.052417}
}

@article{Butt_NatComms_2026,
  title={Demonstration of measurement-free universal logical quantum computation},
  author={Butt, Friederike and Pogorelov, Ivan and Freund, Robert and Steiner, Alex and Meyer, Marcel and Monz, Thomas and others},
  journal={Nat. Commun.},
  year={2026},
  doi = {10.1038/s41467-026-68533-x},
  URL = {https://www.nature.com/articles/s41467-026-68533-x},
  publisher={Nature Publishing Group UK London}
}

\appendix
\onecolumngrid

\section{Derivation of the Two-Level Adiabatic Hamiltonian} \label{Correlation_btw_GS_Leakage_and_E_01} \noindent
Here we present the detailed derivation of the adiabatic Hamiltonian in Eq.~\eqref{H_new_eq}, based on Theorem 1 of Ref.~\cite{Jansen_JMPhys_2007}. This theorem states that the upper bound on the population leakage from the ground state scales as $\varepsilon_{P_{E_0}}(s) = O(\tau^{-2})$, where $s \equiv t/\tau$ denotes the normalized evolution time and $\tau$ represents the characteristic time scale of the adiabatic evolution.

Let $W(s) \equiv \ket{E_0(s)}\bra{E_0(s)}$ denotes the projection onto the ground state. The adiabatic Hamiltonian can be defined as
\begin{equation}
H_{\tau}^{A}(s)\equiv H(s) - \frac{i}{\tau} \left[ \dot{W}(s), W(s) \right]. \label{H_Adiabatic_eq}
\end{equation}

Based on the energy spectrum $\left\{ E_j \right\}$ of the Rydberg Hamiltonian $H_{\text{Ry}}$ in Eq.~\eqref{H_Ry}, $H$ can be writte as
\begin{equation}
H(s)= \sum_{j} E_j (s) \ket{E_j (s)}\bra{E_j (s)} \overset{\mathrm{Diag}}{=} E_1(s) - \Delta E_{01}(s) \cdot W(s), \label{H_eq2} 
\end{equation}
where $\overset{\mathrm{Diag}}{=}$ indicates that the left- and right-hand sides differ only by a scalar multiple of the identity operator.

From the Eqs.~\eqref{H_Adiabatic_eq} and~\eqref{H_eq2},
\begin{subequations} \label{H_Adiabatic_eq2}
\begin{eqnarray}
H_{\tau}^{A}(s) &=& \sum_{j} E_j (s) \ket{E_j (s)}\bra{E_j (s)} - \frac{i}{\tau} \left[ \frac{d}{ds}\ket{E_0(s)}\bra{E_0(s)} , \ket{E_0(s)}\bra{E_0(s)} \right] \label{H_Adiabatic_eq2_1} \\
&=& \sum_{j} E_j (s) \ket{E_j (s)}\bra{E_j (s)} - \frac{i}{\tau} \left[ \frac{d}{ds}\ket{E_0(s)} \cdot \bra{E_0(s)} - \ket{E_0(s)} \cdot \frac{d}{ds}\bra{E_0(s)} \right] \label{H_Adiabatic_eq2_2} \\
&=& \sum_{j} E_j (t) \ket{E_j (t)}\bra{E_j (t)} - i \left[ \frac{d}{dt}\ket{E_0(t)} \cdot \bra{E_0(t)} - \ket{E_0(t)} \cdot \frac{d}{dt}\bra{E_0(t)} \right] \label{H_Adiabatic_eq2_3} \\
&\approx& E_0 (t) \ket{E_0 (t)}\bra{E_0 (t)} + E_1 (t) \ket{E_1 (t)}\bra{E_1 (t)} - i \left[ \frac{d}{dt}\ket{E_0(t)} \cdot \bra{E_0(t)} - \ket{E_0(t)} \cdot \frac{d}{dt}\bra{E_0(t)} \right], \label{H_Adiabatic_eq2_4}
\end{eqnarray}
\end{subequations}
and 
\begin{subequations} \label{dH_dt_eq}
\begin{eqnarray} 
\dot{H}(s) &=&\sum_{j} \biggl[ \dot{E}_j (s) \ket{E_j (s)}\bra{E_j (s)} + E_j (s) \cdot \biggl\{ \frac{d}{ds} \ket{E_j (s)} \cdot \bra{E_j (s)} + \ket{E_j (s)} \cdot \frac{d}{ds} \bra{E_j (s)} \biggr\} \biggr] \label{dH_dt_eq1} \\
&\approx& \dot{E}_1(s) - \frac{d}{ds} \left[ \Delta {E}_{01}(s) \right] \cdot W(s) - \Delta E_{01} (s) \cdot \dot{W}(s). \label{dH_dt_eq2}
\end{eqnarray}
\end{subequations}

From the Eqs.~\eqref{H_eq2},~\eqref{H_Adiabatic_eq2} and~\eqref{dH_dt_eq},
\begin{subequations} \label{H_Adiabatic_eq3}
\begin{eqnarray}
H_{\tau}^{A}(s=t/\tau) &\approx& E_0 (t) \ket{E_0 (t)}\bra{E_0 (t)} + E_1 (t) \ket{E_1 (t)}\bra{E_1 (t)} \nonumber \\ 
&& - \frac{i}{\tau \Delta E_{01} (s)} \left[ \dot{E}_1(s) - \frac{d}{ds} \left[ \Delta {E}_{01}(s) \right] \cdot W(s) - \dot{H}(s) , W(s) \right] \label{H_Adiabatic_eq3_1} \\
&=& E_1 (s) - \Delta E_{01} (s) \cdot W(s) + \frac{i}{\tau \Delta E_{01} (s)} \left[ \dot{H}(s),  W(s) \right] \label{H_Adiabatic_eq3_3} 
\end{eqnarray}
\end{subequations} 

Comparing the Eqs.~\eqref{H_Adiabatic_eq},~\eqref{H_eq2} and~\eqref{H_Adiabatic_eq3_3}, 
\begin{equation}
W(s)\overset{\mathrm{Diag}}{=} -\frac{H(s)}{\Delta E_{01}(s)}, \qquad \dot{W}(s) \overset{\left[W(s), \cdot \right]}{=} \frac{\dot{H}(s)}{\Delta E_{01}(s)}, 
\end{equation}
where $\overset{\left[W(s), \cdot \right]}{=}$ indicates that the two sides are equivalent when evaluated within the commutator $\left[W(s), \cdot \right]$.
This leads to $H_{\rm ad}$ (Eq.~\eqref{H_new_eq}): 
\begin{equation} \label{H_new_eq2}
H_{\tau}^{A}(s=t/\tau) \overset{\mathrm{Diag}}{=} \frac{ \bra{E_1} \frac{dH}{dt} \ket{E_0} }{ \Delta E_{01} } s_y - \frac{ \Delta E_{01} }{2} s_z \, \left[ \, \equiv H_{\rm ad}(t) \, \right].
\end{equation}


\section{Atomic Coordinates for the Quasi-One-Dimensional Chains and Two-Dimensional Triangular Lattices} \label{Atom_Coords_for_Q1D_Chains_and_2D_Tri_Lattices}

\begin{table*}[th!]
\caption{Atom positions of $Q_{1D, 10}$}
\label{Table_Atom_Coords_of_Q1D_Chains}
\begin{center}
\begin{tabular}{@{}crlrlrl@{}}
\hline \hline
Graph & & \vline & \multicolumn{4}{c}{Atom positions $(x,y)$ ($\mu$m)} \\
\hline
\multirow{7}{*}{$Q_{1D, 10}$} & & \vline & & & & \\
& & \vline & $x_1$: & (6.93, 8.00) & $x_2$:& (20.78, 8.00) \\
& & \vline & $x_3$: & (34.64, 8.00) &  $x_4$:& (0.00, 4.00) \\
& & \vline & $x_5$: & (13.85, 4.00)  & $x_6$:& (27.71, 4.00)  \\
& & \vline & $x_7$: & (41.57, 4.00) & $x_8$:& (6.93, 0.00) \\
& & \vline & $x_9$: & (20.78, 0.00) & $x_{10}$:& (34.64, 0.00) \\
& & \vline &  &  &  &  \\
\hline \hline
\end{tabular}
\end{center}
\end{table*}

\begin{table*}[th!]
\caption{Atom positions of $T_{D, 25}$, $T_{H, 37}$}
\label{Table_Atom_Coords_of_Tri_2D_Lattices}
\begin{center}
\begin{tabular}{@{}crlrlrlccrlrlrl@{}}
\hline \hline
Graph & & \vline &  \multicolumn{4}{c}{Atom positions $(x,y)$ ($\mu$m)} & \vline & Graph & & \vline & \multicolumn{4}{c}{Atom positions $(x,y)$ ($\mu$m)} \\
\hline
\multirow{21}{*}{$T_{D, 25}$} & & \vline  &  &  &  &  & \vline & \multirow{21}{*}{$T_{H, 37}$} & & \vline  &  &  &   \\
& & \vline  &  &  &  &  & \vline & & & \vline & $x_1$: & (20.78, 0.00) & $x_2$:& (13.86, 4.00) \\
& & \vline  &  &  &  &  & \vline & & & \vline & $x_3$: & (27.71, 4.00) &  $x_4$:& (6.93, 8.00) \\
& & \vline  &  &  &  &  & \vline & & & \vline & $x_5$: & (20.78, 8.00)  & $x_6$:& (34.64, 8.00) \\
& & \vline & $x_1$: & (27.71, 4.00) & $x_2$:& (20.78, 8.00) & \vline & & & \vline & $x_7$: & (0.00, 12.00) & $x_8$:& (13.86, 12.00) \\
& & \vline & $x_3$: & (34.64, 8.00) &  $x_4$:& (13.86, 12.00) & \vline & & & \vline & $x_9$: & (27.71, 12.00) & $x_{10}$:& (41.57, 12.00) \\ 
& & \vline & $x_5$: & (27.71, 12.00)  & $x_6$:& (41.57, 12.00) & \vline & & & \vline  & $x_{11}$: & (6.93, 16.00) & $x_{12}$:& (20.78, 16.00) \\
& & \vline & $x_7$: & (6.93, 16.00) & $x_8$:& (20.78, 16.00) & \vline & & & \vline & $x_{13}$: & (34.64, 16.00) & $x_{14}$:& (0.00, 20.00) \\
& & \vline & $x_9$: & (34.64, 16.00) & $x_{10}$:& (48.50, 16.00) & \vline & & & \vline & $x_{15}$: & (13.86, 20.00) & $x_{16}$:& (27.71, 20.00) \\
& & \vline & $x_{11}$: & (0, 20.00) & $x_{12}$:& (13.86, 20.00) & \vline & & & \vline & $x_{17}$: & (41.57, 20.00) & $x_{18}$:& (6.93, 24.00) \\
& & \vline & $x_{13}$: & (27.71, 20.00) & $x_{14}$:& (41.57, 20.00) & \vline & & & \vline & $x_{19}$: & (20.78, 24.00) & $x_{20}$:& (34.64, 24.00) \\
& & \vline & $x_{15}$: & (55.43, 20.00) & $x_{16}$:& (6.93, 24.00) & \vline & & & \vline & $x_{21}$: & (0.00, 28.00) & $x_{22}$:& (13.86, 28.00) \\
& & \vline & $x_{17}$: & (20.78, 24.00) & $x_{18}$:& (34.64, 24.00) & \vline & & & \vline & $x_{23}$: & (27.71, 28.00) & $x_{24}$:& (41.57, 28.00) \\
& & \vline & $x_{19}$: & (48.50, 24.00) & $x_{20}$:& (13.86, 28.00) & \vline & & & \vline & $x_{25}$: & (6.93, 32.00) & $x_{26}$:& (20.78, 32.00) \\
& & \vline & $x_{21}$: & (27.71, 28.00) & $x_{22}$:& (41.57, 28.00) & \vline & & & \vline & $x_{27}$: & (34.64, 32.00) & $x_{28}$:& (0.00, 36.00) \\
& & \vline & $x_{23}$: & (20.78, 32.00) & $x_{24}$:& (34.64, 32.00) & \vline & & & \vline & $x_{29}$: & (13.86, 36.00) & $x_{30}$:& (27.71, 36.00) \\
& & \vline & $x_{25}$: & (27.71, 36.00) & & & \vline & & & \vline & $x_{31}$: & (41.57, 36.00) & $x_{32}$:& (6.93, 40.00) \\
& & \vline  &  &  &  &  & \vline & & & \vline & $x_{33}$: & (20.78, 40.00) & $x_{34}$:& (34.64, 40.00) \\
& & \vline  &  &  &  &  & \vline & & & \vline & $x_{35}$: & (13.86, 44.00) & $x_{36}$:& (27.71, 44.00) \\
& & \vline  &  &  &  &  & \vline & & & \vline & $x_{37}$: & (20.78, 48.00) &  \\
& & \vline  &  &  &  &  & \vline & & & \vline  &  &  &  \\
\hline \hline
\end{tabular}
\end{center}
\end{table*}

\begin{table*}[th!]
\caption{Atom positions of $Q'_{1D, 23}$}
\label{Table_Atom_Coords_of_Q1D_Chain_Extension_N23}
\begin{center}
\begin{tabular}{@{}crlrlrl@{}}
\hline \hline
Graph & & \vline & \multicolumn{4}{c}{Atom positions $(x,y)$ ($\mu$m)} \\
\hline
\multirow{14}{*}{$Q'_{1D, 23}$} & & \vline & & & & \\
& & \vline & $x_1$: & (13.86, 4.00) & $x_2$:& (20.78, 0.00) \\
& & \vline & $x_3$: & (27.71, 4.00) &  $x_4$:& (34.64, 0.00) \\
& & \vline & $x_5$: & (41.57, 4.00)  & $x_6$:& (48.50, 0.00)  \\
& & \vline & $x_7$: & (55.43, 4.00) & $x_8$:& (6.93, 8.00) \\
& & \vline & $x_9$: & (20.78, 8.00) & $x_{10}$:& (34.64, 8.00) \\
& & \vline & $x_{11}$: & (48.50, 8.00) & $x_{12}$:& (6.93, 16.00) \\
& & \vline & $x_{13}$: & (6.93, 24.00) & $x_{14}$:& (13.86, 28.00) \\
& & \vline & $x_{15}$: & (20.78, 24.00) & $x_{16}$:& (27.71, 28.00) \\
& & \vline & $x_{17}$: & (34.64, 24.00) & $x_{18}$:& (41.57, 28.00) \\
& & \vline & $x_{19}$: & (48.50, 24.00) & $x_{20}$:& (55.43, 28.00) \\
& & \vline & $x_{21}$: & (20.78, 32.00) & $x_{22}$:& (34.64, 32.00) \\
& & \vline & $x_{23}$: & (48.50, 32.00) &  &   \\
& & \vline &  &  &  &  \\
\hline \hline
\end{tabular}
\end{center}
\end{table*}

\twocolumngrid

\end{document}